\documentclass[
]{ceurart}

\usepackage{listings}
\usepackage[defaultlines=2,all]{nowidow}
\begin{document}

\copyrightyear{2024}
\copyrightclause{Copyright for this paper by its authors.
  Use permitted under Creative Commons License Attribution 4.0
  International (CC BY 4.0).}

\conference{EKAW 2024: EKAW 2024 Workshops, Tutorials, Posters and Demos, 24th International Conference on Knowledge Engineering and Knowledge Management (EKAW 2024), November 26-28, 2024, Amsterdam, The Netherlands.}

\title{Realizing a Collaborative RDF Benchmark Suite in Practice}

\author[1]{Piotr Sowiński}[%
orcid=0000-0002-2543-9461,
email=piotr.sowinski.dokt@pw.edu.pl,
]
\cormark[1]
\address[1]{Warsaw University of Technology, Pl. Politechniki 1, 00-661 Warsaw, Poland}

\author[1]{Maria Ganzha}[%
orcid=0000-0001-7714-4844,
email=maria.ganzha@pw.edu.pl,
]

\cortext[1]{Corresponding author.}

\begin{abstract}
Collaborative mechanisms allow benchmarks to be updated continuously and adjust to the changing requirements and new use cases. This paradigm is employed for example in the field of machine learning, but up until now there were no examples of truly open and collaborative benchmarks for RDF systems. In this demo paper we present the collaboration functionalities of RiverBench, an open, multi-task RDF benchmark suite. Owing to its fully open and community-driven design, RiverBench allows any researcher or practitioner to submit a new dataset or benchmark task, report performed benchmark runs, and edit any resource in the suite. RiverBench's collaboration system is itself based on RDF and Linked Data mechanisms, and every resource in the suite has machine-readable RDF metadata. The showcased functionalities together make up a first-of-a-kind fully open and collaborative RDF benchmark suite. These features are meant to encourage other researchers to contribute to RiverBench, and make it a long-term project sustained by the community.
\end{abstract}


\maketitle

\section{Introduction}

Benchmarking Resource Description Framework (RDF) systems is a popular topic in the Semantic Web and Knowledge Graph research communities. Benchmarks not only allow one to compare how different systems perform in certain scenarios, but they also help drive innovation, by identifying gaps in performance and functionalities. Most often, a benchmark task is defined in a research paper, with the dataset(s) and benchmark code attached as supplementary resources. However, it is very hard for such benchmarks to evolve over time, for example, by adding new datasets or metrics contributed by the community. Furthermore, the dispersed nature of these benchmarks makes it  difficult to follow consistent standards in benchmark specification, dataset distribution, and result reporting. This in turn reflects negatively upon benchmark reproducibility and slows down progress in the field. We postulate that these issues could be at least partially overcome with an \textbf{open, collaborative benchmark suite} which could naturally evolve over time to suit the changing needs of the community, while maintaining consistent standards.

Although many benchmarks and benchmark suites for RDF systems were proposed, few of them incorporate any collaborative aspects. The most prominent example here is the Linked Data Benchmark Council (LDBC), which maintains several graph benchmarks, including RDF-based workloads (e.g., the Social Network Benchmark~\cite{erling2015ldbc}). Organizations and individuals can contribute to LDBC with a paid membership and participation in its working groups. LDBC also gathers audited benchmark results, but this process is also paid and generally reserved to large industry actors. Although LDBC benchmarks are technically collaborative, their main target are production-ready commercial systems, and the contribution process is complex. The only other example that we could find in the RDF community is the now-defunct Liquid Benchmarks platform~\cite{sakr2011liquid}. It was a collaborative cloud application for running and reporting benchmark results in a variety of tasks, including SPARQL query processing. However, as the platform is not available anymore, it is difficult to assess what was the exact extent of its collaborative capabilities.

The idea of collaborative benchmarks is more popular in other areas of computer science, especially in machine learning (ML) and natural language processing (NLP). For example, the GLUE benchmark for natural language understanding~\cite{wang2018glue} allows researchers to submit their benchmark results via a form on the benchmark website. However, this only works for a fixed set of tasks defined in GLUE. The NLP-Progress website~\cite{ruder2024nlp} is a more open-ended project, where researchers are invited to report their NLP benchmark results via pull requests on GitHub. The site then displays these results in the form of HTML tables grouped by benchmark task and dataset, with no machine-readable content. The BenchOpt benchmark suite for ML and optimization~\cite{moreau2022benchopt} goes a step further, by being tightly integrated with a dedicated benchmark runner. The runner can semi-automatically report benchmark results (via a GitHub pull request) to be displayed on the website. Other contributions (e.g., to the benchmark code) are also welcome and can be submitted using pull requests.

\paragraph{Contribution.} In this demo paper we showcase the recently-developed collaborative features of RiverBench~\cite{sowinski2023riverbench}, an open, multi-task benchmark suite for RDF systems. RiverBench is designed to be fully open, transparent, and community-driven from the start. It allows submitting new benchmark tasks and datasets, reporting benchmark results, and editing any other aspect of the suite. The collaboration architecture is based on RDF metadata and Linked Data mechanisms, making every resource in RiverBench machine-accessible. The presented approach goes beyond the current state of the art by being (to the best of our knowledge) the first RDF benchmark suite with such extensive open collaboration functionalities. It is also a major technical advancement over the aforementioned benchmark projects in other fields of computer science, which do not fully employ the FAIR principles, machine-readable RDF metadata, or the Linked Data mechanisms.

\section{Collaborative Architecture of RiverBench}

RiverBench collects both benchmark tasks and datasets. Although RiverBench focuses on streaming use cases (e.g., streaming RDF graphs over the network), it also includes non-streaming tasks, such as loading RDF data into a triple store, or serializing an RDF graph to bytes. Benchmark profiles group datasets with shared technical characteristics (e.g., datasets with RDF triples only, no quads), to make dataset selection easier. Every resource in RiverBench has an HTML documentation page and machine-readable RDF metadata, available via the content negotiation mechanism. The suite makes extensive use of Continuous Integration / Continuous Deployment (CI/CD) scripts and GitHub's free infrastructure to automate most tasks.

\begin{figure}[htb]
    \centering
    \includegraphics[width=\linewidth]{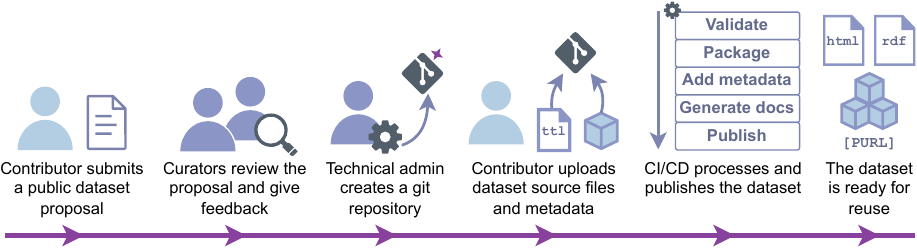}
    \caption{The process of proposing and publishing a new dataset in RiverBench.}
    \label{fig:dataset-process}
\end{figure}

Benchmark datasets are added to the suite through a public community process, visualized in Figure~\ref{fig:dataset-process}. Any practitioner or researcher is invited to submit their dataset, as long as it meets basic technical requirements and can be published under a permissive license. Firstly, the contributor fills out the dataset proposal form on GitHub, giving details about the dataset's source, license, original use case, etc. The proposal is then reviewed by RiverBench curators, who check if the dataset meets the publicly-available criteria for new datasets (e.g., open license, properly specified authorship, sufficient size, clear use case) and if it would be a valuable addition. The curators are members of the RDF benchmarking community, who besides reviewing dataset proposals also have a deciding voice in major decisions shaping RiverBench. As of November 2024, there are four registered curators from three different institutions\footnote{\url{https://w3id.org/riverbench/v/dev/documentation/maintainers}}. After the proposal is accepted by the curators, a technical administrator creates a new GitHub repository for the dataset, and instructs the contributor on how to proceed. The contributor then uploads the dataset for processing and fills out a metadata file in the Turtle language, describing the dataset in a structured manner. The dataset and the metadata are then automatically processed by CI/CD: validating its contents, re-packaging the dataset in several formats, adding more metadata, generating documentation pages, and publishing it on the website under a permanent URL (PURL).

% if we have room, a diagram for proposing a new task

% or a screenshot of a dataset page...

Proposing new benchmark tasks follows a very similar public community process. The contributor also fills out a proposal form on GitHub, which is reviewed by the curators who check if the task meets the pre-defined criteria. Afterwards, the contributor is asked to write down the description and metadata of the task in a Turtle file. The new task is then automatically processed by CI/CD (validation, packaging, documentation generation, publishing).

\begin{figure}[htb]
    \centering
    \includegraphics[width=\linewidth]{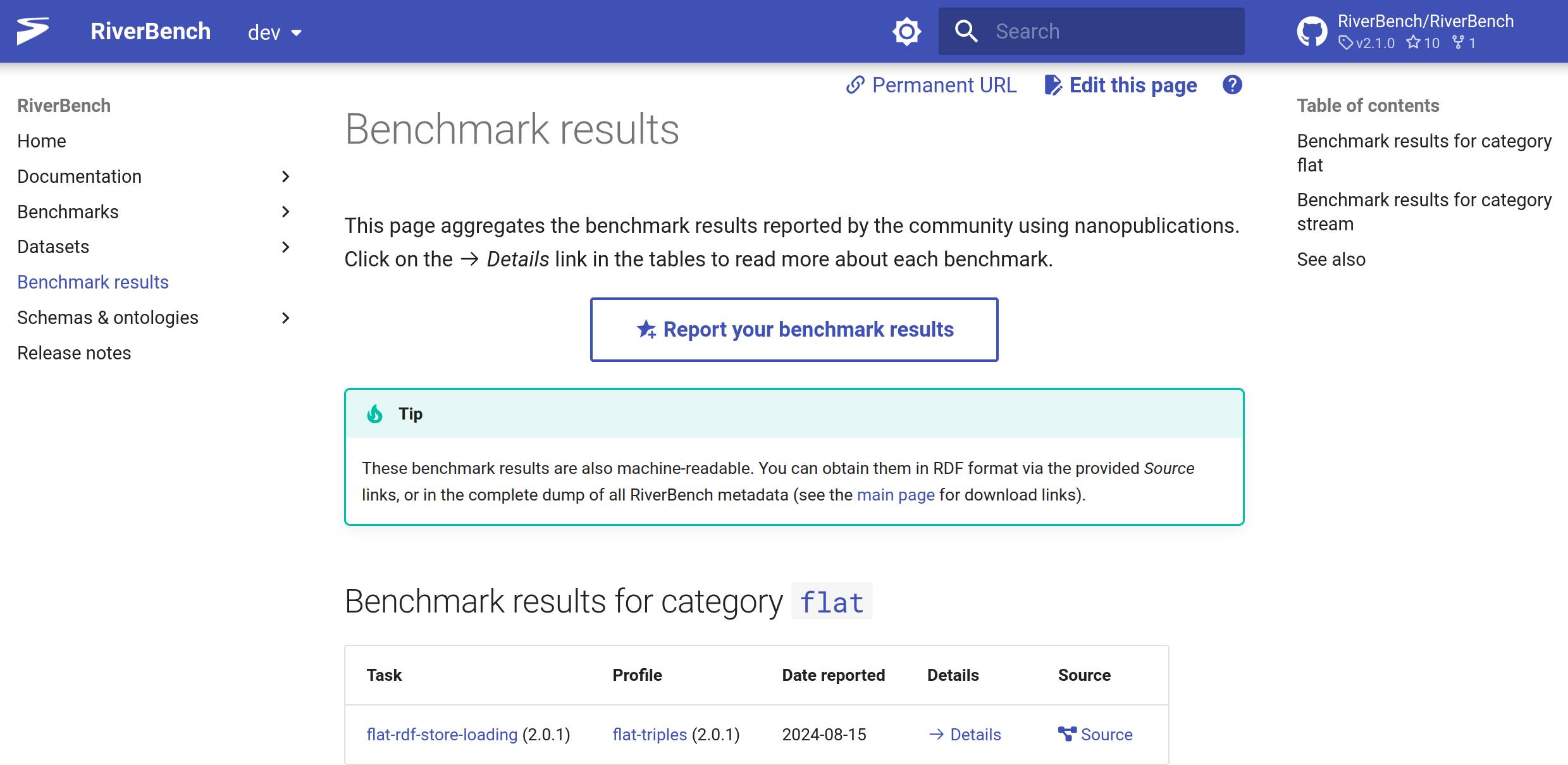}
    \caption{Screenshot from the RiverBench benchmark results page (\url{https://w3id.org/riverbench/v/dev/results}).}
    \label{fig:rb-results-page}
\end{figure}

The source code of tasks, datasets, profiles (groups of datasets), documentation pages, and other resources of RiverBench are hosted on GitHub as Markdown or Turtle files in one of the several repositories. To make editing these resources as easy as possible, every page on the RiverBench website has an ``Edit this page'' button (visible at the top of Figure~\ref{fig:rb-results-page}). The button immediately redirects the user to the correct file on GitHub, where they can edit it and submit a pull request with the changes. After the pull request is validated by a CI job and accepted by a curator or technical administrator, the changes automatically appear on the website via a CI/CD pipeline.

\begin{figure}[htb]
    \centering
    \includegraphics[width=\linewidth]{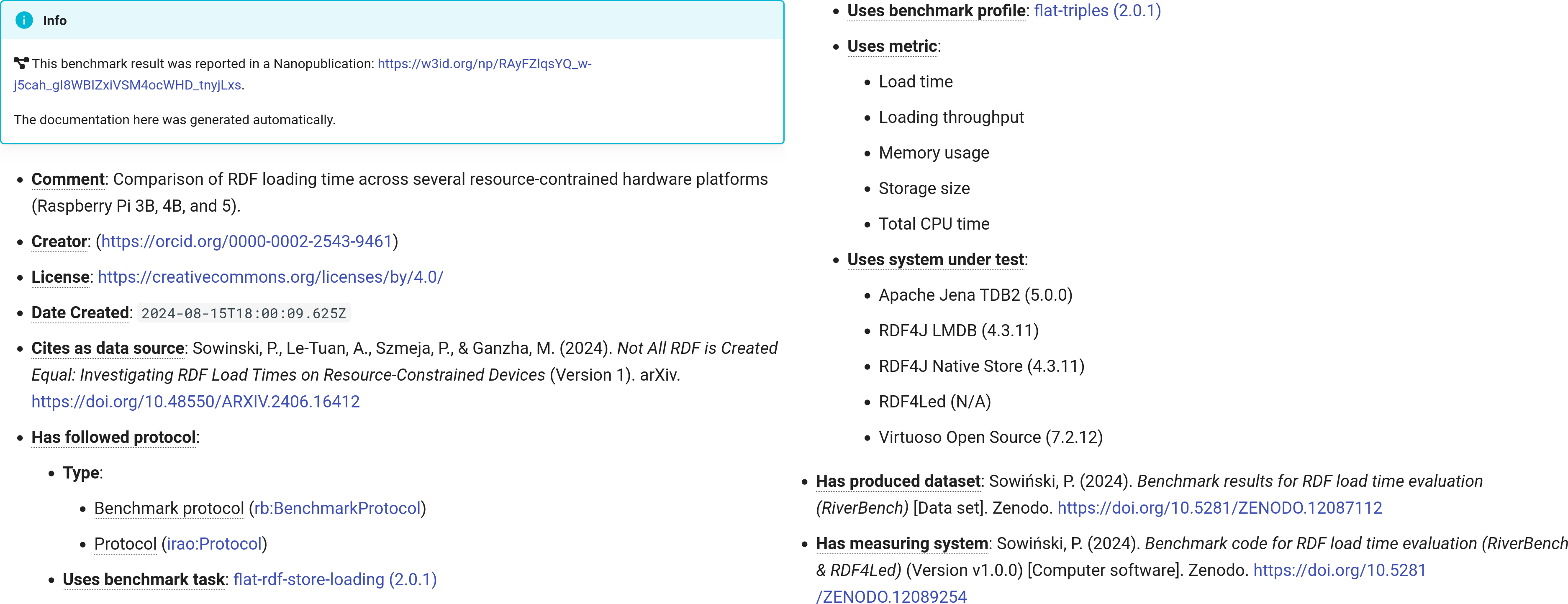}
    \caption{Example benchmark run report details taken from a nanopublication, as rendered on the RiverBench website. The screenshot was modified into a two-column layout for visual clarity.}
    \label{fig:result-example}
\end{figure}

RiverBench also collects benchmark run reports and displays them on the website (Figures~\ref{fig:rb-results-page} and~\ref{fig:result-example}). Each benchmark run report is a nanopublication~\cite{kuhn2013broadening} -- a small unit of scientific knowledge encoded as an RDF dataset. These nanopublications use the RiverBench ontology and the Informatics Research Artifact Ontology (IRAO)~\cite{nguyen2021ontology} to structure the information on the used RiverBench task, profile (group of datasets), benchmark code, evaluated systems, and more.

\begin{figure}[htb]
    \centering
    \includegraphics[width=1.0\linewidth]{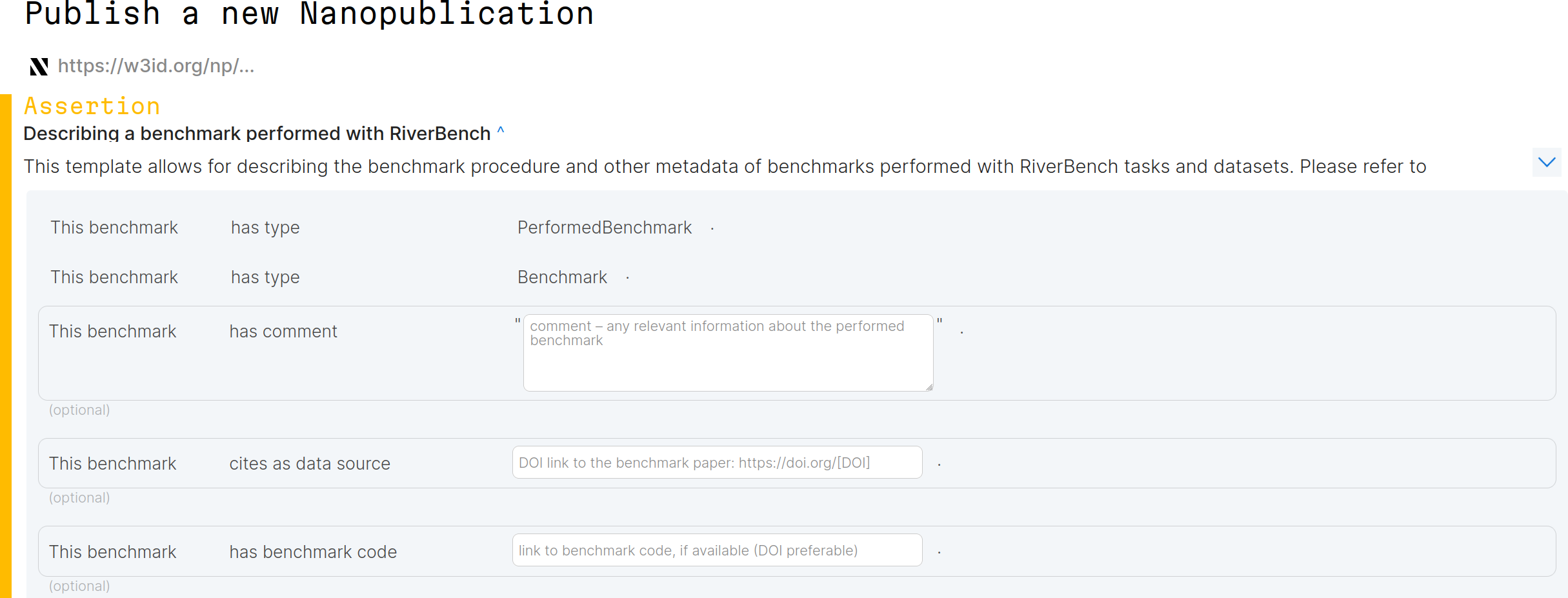}
    \caption{Screenshot from Nanodash~\cite{kuhn2021semantic} (AGPL-3.0 license) showing a fragment of the form used to submit benchmark results in RiverBench.}
    \label{fig:nanodash}
\end{figure}

Nanopublications with benchmark run reports can be submitted by any researcher with an ORCID account to the decentralized nanopublication network. A nanopublication template for the web application Nanodash~\cite{kuhn2021semantic} was prepared, which simplifies the process, requiring the user to only fill out a web form (Figure~\ref{fig:nanodash}). After the nanopublication is submitted, it will be automatically discovered by RiverBench's CI/CD. It is then turned into human-readable documentation on the suite's website (Figure~\ref{fig:result-example}), and republished in metadata dumps for easier reuse. This workflow is similar to a mechanism proposed by the RDF Stream Taxonomy~\cite{sowinski2024rdf}, where nanopublications are used to create a ``living literature review'' of RDF streaming.

\section{Demonstration}

In the live demonstration, we will use the public RiverBench website\footnote{\url{https://w3id.org/riverbench}} to briefly showcase the main collaboration functionalities of RiverBench. The documentation pages linked in the footnotes contain more details about each functionality.

\textbf{Reporting benchmark results.} We will show how to report the results of benchmarks conducted using RiverBench\footnote{\url{https://w3id.org/riverbench/v/dev/documentation/reporting-results}}. First, we will fill out the benchmark result form in Nanodash~\cite{kuhn2021semantic} with example data and submit it. Then, we will show how it appears on the website and in the semantic metadata of RiverBench.

\textbf{Editing documentation and metadata.} We will briefly explain the mechanism of editing documentation and the locations of edit buttons on the RiverBench website\footnote{\url{https://w3id.org/riverbench/v/dev/documentation/editing-docs}}. Then, we will demonstrate editing a documentation page and the metadata of a resource (e.g., task or dataset). We will then show to the participants the changes as they appear on the RiverBench website.

\textbf{Proposing new datasets.} We will briefly explain the dataset proposal process\footnote{\url{https://w3id.org/riverbench/v/dev/documentation/creating-new-dataset}}, and show how to fill out a dataset proposal form and the metadata file.

\textbf{Proposing new benchmark tasks.} Following a similar structure as with dataset proposals, we will explain the task proposal process\footnote{\url{https://w3id.org/riverbench/v/dev/documentation/creating-new-task}}. We will then show how to fill out the form and the metadata file.

\section{Conclusion}

In this demo paper, we showcased the collaborative features of RiverBench, an open RDF benchmark suite. The presented four functionalities (reporting benchmark results, editing existing resources, submitting datasets, and proposing benchmark tasks) are supported by comprehensive automation (CI/CD). Additionally, RiverBench itself is a use case of knowledge management techniques, as it extensively uses RDF metadata and Linked Data mechanisms, making this contribution doubly interesting and relevant to the Semantic Web community. The features were designed to be easy to use, open to any contributor, and rely on public and transparent community processes.

To the best of our knowledge, RiverBench is the first RDF benchmark suite to implement collaborative features so widely. We hope that these easy-to-use functionalities will encourage the Semantic Web and Knowledge Graph communities to contribute to RiverBench and join the project as curators, ensuring its long-term relevance and usefulness. The interest of the community ultimately depends on whether the project fulfills an important need or not -- in this case, if it useful for benchmarking RDF systems. Currently, RiverBench can be used with nine diverse, scientifically relevant benchmark tasks, but this is expected to be expanded in the future. Based on the early feedback from other researchers, we plan to expand the suite's infrastructure to support more use cases and tasks. We will also continue improving the suite's documentation and automation system, to match the changing expectations of the community.

In contrast to the now-defunct Liquid Benchmarks platform~\cite{sakr2011liquid}, RiverBench is hosted on permanently free infrastructure (GitHub, w3id.org, and the nanopublication network), which should ensure its longevity. While w3id.org and the nanopublication network are maintained by multiple independent actors and therefore should be stable, GitHub could conceivably cease its operations in the future, or stop providing free services. RiverBench is prepared for that eventuality, as all its resources are hosted under permanent URLs, which can be easily redirected to a different service provider. Because the core features of the suite only require a static file hosting server, the minimal migration would not be problematic. The only code is executed as part of CI/CD pipelines, which could be ported to alternative services (e.g., GitLab).

\section*{Online Resources}

\begin{itemize}
    \item \textbf{RiverBench website} -- the subject of the demo: \url{https://w3id.org/riverbench}
    \item \textbf{RiverBench documentation}: \url{https://w3id.org/riverbench/v/dev/documentation}
    \item \textbf{Source code on GitHub}: \url{https://github.com/RiverBench}
\end{itemize}

\bibliography{bibliography}

\end{document}